\documentclass[11pt,onecolumn]{article}

\usepackage{bm}
\usepackage{amsfonts}
\usepackage{amsmath}
\usepackage{amssymb}
\usepackage{cite}
\usepackage[amsmath,thmmarks]{ntheorem}

\usepackage{color, soul}
\usepackage{epic, eepic}
\usepackage[top=1in, bottom=1in, left=1.25in, right=1.25in]{geometry}
\usepackage{appendix}

\usepackage{theorem}

\theoremheaderfont{\sc}\theorembodyfont{\upshape}
\theoremstyle{nonumberplain}
\theoremseparator{}
\theoremsymbol{\rule{1ex}{1ex}}
\newtheorem{proof}{Proof}

\newcommand{\red}{\textcolor[rgb]{0.00,0.00,0.00}}

\begin{document}

\title{Perturbation Analysis of Orthogonal Matching Pursuit}

\author{Jie~Ding,~Laming~Chen,~and~Yuantao~Gu\thanks{This work was partially
supported by National Natural Science Foundation of
China (NSFC 60872087 and NSFC U0835003) and Agilent Technologies
Foundation \# 2205. The authors are with Department of
Electronic Engineering, Tsinghua University, Beijing 100084, China. The corresponding author of this paper is Yuantao Gu (gyt@tsinghua.edu.cn).}}

\date{Received June 07, 2011; accepted Sep. 05, 2012.\\\vspace{1em}
This article appears in \textsl{IEEE Transactions on Signal Processing}, 61(2): 398-410, 2013.}

\maketitle

\begin{abstract}
Orthogonal Matching Pursuit (OMP) is a canonical greedy pursuit algorithm for sparse approximation.
Previous studies of OMP have considered the recovery of a sparse signal $\bm x$ through $\bm \Phi$
and $\bm y=\bm \Phi \bm x+\bm b$, where $\bm \Phi$ is a matrix with more columns than rows and $\bm
b$ denotes the measurement noise. In this paper, based on Restricted Isometry Property (RIP), the
performance of OMP is analyzed under general perturbations, which means both $\bm y$ and $\bm \Phi$
are perturbed. Though the exact recovery of an almost sparse signal $\bm x$ is no longer feasible,
the main contribution reveals that the support set of the best $k$-term approximation of $\bm x$
can be recovered under reasonable conditions. The error bound between $\bm x$ and the estimation of
OMP is also derived. By constructing an example it is also demonstrated that the sufficient
conditions for support recovery of the best $k$-term approximation of $\bm x$ are rather tight.
When $\bm x$ is strong-decaying, it is proved that the sufficient conditions for support recovery
of the best $k$-term approximation of $\bm x$ can be relaxed, and the support can even be recovered
in the order of the entries' magnitude. Our results are also compared in detail with some related previous ones.

\end{abstract}

\section{Introduction}

Finding the sparse solution of an underdetermined linear equation
\begin{equation}\label{dj12}
\bm y=\bm \Phi \bm  x
\end{equation}
is one of the basic problems in some fields of signal processing, where $\bm y \in \mathbb{C}^m$
and $\bm \Phi\in \mathbb{C}^{m\times n}$ with $m<n$. The basic problem (\ref{dj12}) has arisen in
many applications, including Sparse Component Analysis (SCA) \cite{Gribonval,Cichocki} and Blind
Source Separation (BSS) \cite{Bofill,Georgiev}. Since the introduction of Compressed Sensing (CS)
\cite{Donoho,Romberg,Compressive,Baraniuk}, the problem (\ref{dj12}) has received significant
attention in the past decade. In the field of CS, $\bm y$ denotes the measurement vector, $\bm
\Phi$ is called the sensing matrix, and $\bm x$ is the sparse or almost sparse signal to be
recovered.

Various algorithms have been proposed to recover $\bm x$. They roughly fall into two categories.

\textbf{Convex relaxation:} Based on linear programming technique, finding the sparse solution to
(\ref{dj12}) can be relaxed to a convex optimization problem, also known as Basis Pursuit (BP)
\cite{Romberg}. As for the case of noisy measurements, the problems of Least Absolutely Shrinkage
and Selection Operator (LASSO) \cite{LASSO} and Basis Pursuit De-Noising (BPDN) \cite{BPDN} are
introduced. Algorithms used to complete the convex optimization include Interior-point Methods
\cite{Kim}, Projected Gradient Methods \cite{Figueiredo}, and Iterative Thresholding
\cite{Daubechies}.

\textbf{Greedy pursuits:} Most of these algorithms build up an approximated set of nonzero
locations by making locally optimal choices in each iteration. Several popular ones are Orthogonal
Matching Pursuit (OMP) \cite{Schen,origin,Mallat}, Regularized Orthogonal Matching Pursuit (ROMP)
\cite{Deanna}, Compressive Sampling Matching Pursuit (CoSaMP) \cite{Needell}, Subspace Pursuit (SP)
\cite{Wei}, and Iterative Hard Thresholding (IHT) \cite{Thomas}.

For the scenario of no noise or perturbation, the recovery process can be formulated as
\begin{equation*}
(N_0) \qquad \hat{\bm x}=R(\bm y, \bm \Phi, \cdots),
\end{equation*}
where $R(\cdot)$ denotes the process of a recovery algorithm, with the inputs listed in the
following brackets, and $\hat{\bm x}$ denotes the output (i.e. the approximation of the original
sparse signal $\bm x$). Process of $(N_0)$ is non-perturbed, thus the sparse signal can be exactly
recovered under suitable conditions. For example, under certain conditions, BP \cite{Candes,Cai},
OMP \cite{Tropp,Joel,Mark,Entao,review2,Brownian,remark}, ROMP \cite{Deanna}, CoSaMP \cite{Needell}
and SP \cite{Wei} all guarantee exact recovery of $\bm x$.

In practical applications, the measurement vector $\bm y$ is often contaminated by noise. Thus a
perturbed measurement vector in the form of
\begin{equation}
\tilde{\bm y}=\bm y+\bm b
\end{equation}
is considered, where $\bm b$ denotes the measurement noise. In such scenario, the recovery
process can be formulated as
\begin{equation*}
(N_1) \qquad \hat{\bm x}=R(\tilde{\bm y}, \bm \Phi, \cdots).
\end{equation*}
Plentiful studies of recovery algorithms including BP \cite{Candes,BPN1,Guangwu,Emmanuel,uOMP}, OMP
\cite{review2,Brownian,BPN1,uOMP,Denis}, ROMP \cite{Roman}, CoSaMP \cite{Needell}, SP \cite{Wei},
IHT \cite{Thomas}, and Sequential Orthogonal Matching Pursuit (SeqOMP) \cite{Fletcher} have
considered the recovery accuracy in $(N_1)$ process. Define the support set ${\rm supp}(\cdot)$ as
the set composed of the locations of all nonzero entries of a vector. It has been shown that OMP
will exactly recover the support set of a sparse signal $\bm x$ from the perturbed measurement
vector, i.e. ${\rm supp}(\hat{\bm x})={\rm supp}(\bm x)$, if certain requirements are satisfied
with the coherence parameter $\mu$ (Th.5.1 in \cite{BPN1}, Th.3.1 in \cite{Denis}) or Restricted
Isometry Property (RIP) (Th.2 in \cite{review2}). It is worth mentioning that the noise $\bm b$ is
assumed to be deterministic and unknown, and to have a bounded norm (also this paper's setting). In
another common setting, which is not being handled in this paper, $\bm b$ denotes white Gaussian
noise and the recovery of support set for OMP is discussed based on probability (Th.4 in \cite{uOMP}).

Existing results have mainly focused on the measurement noise, yet results considering the general
perturbations are relatively rare. Here, the general perturbations involve a perturbed sensing
matrix as well as a perturbed measurement vector. Two situations are considered in this paper from
different perspective of views.

The first scenario is from user's perspective of view. By measuring an unknown system, one obtains
its sensing matrix which is inaccurate. Thus the sensing process is in the form of
\begin{equation}\label{dj13}
\tilde{\bm y}=\bm \Phi \bm x+\bm b,\ \ \tilde{\bm \Phi}=\bm \Phi+\bm E,
\end{equation}
with recovery process
\begin{equation*}
(N_2) \qquad \hat{\bm x}=R(\tilde{\bm y}, \tilde{\bm \Phi}, \cdots).
\end{equation*}
The system perturbation $\bm E$ is introduced because of mismodeling of the system, or the error
involved during system calibration. Since the available sensing matrix is the perturbed $\tilde{\bm
\Phi}$ instead of $\bm \Phi$, the conditions for recovery are also in terms of the former.

The second scenario is from designer's perspective of view, which means the system perturbation
$\bm E$ is introduced by physical implementation of a designed system model $\bm \Phi$
\cite{Moshe}. Thus the sensing process is in the form of
\begin{equation}\label{r28}
\tilde{\bm y}=\tilde{\bm \Phi}\bm x+\bm b,\ \ \tilde{\bm \Phi}=\bm \Phi+\bm E,
\end{equation}
with recovery process
\begin{equation*}
(N_2') \qquad \hat{\bm x}=R(\tilde{\bm y}, \bm \Phi, \cdots).
\end{equation*}
Since the available sensing matrix is the ideal one, the conditions for recovery should be in terms
of $\bm \Phi$ in this scenario.

Herman and Strohmer have studied the accuracy of BP solution in $(N_2)$ process \cite{Matthew}.
Later, Herman and Needell also gave the recovery error of CoSaMP \cite{mixed}. However, as far as
we know, few works have been done yet on the recovery error or perfect support recovery of OMP
under general perturbations.

Analysis of OMP considering general perturbations and support recovery may benefit the analysis of
other greedy algorithms. In some applications, recovering the support set other than a more
accurate estimation is a fundamental concern (e.g., in the reconstruction stage of the modulated
wideband converter (MWC) \cite{Moshe,boost}). In this paper, a completely perturbed scenario in the
form of (\ref{dj13}) is considered and the performance of OMP in $(N_2)$ process is studied. It is
shown that under certain RIP based conditions, the locations of $k$ largest magnitude entries of an
almost sparse signal $\bm x$ can be exactly recovered via OMP. Furthermore, an upper bound on the
relative recovery error is given. It is also demonstrated that the results generalize the previous
study concerning OMP in $(N_0)$ process in \cite{Mark,Entao,review2,remark}. The completely
perturbed scenario (\ref{r28}) together with $(N_2')$ process is also briefly discussed.

The rest of the paper is organized as follows. Section \uppercase\expandafter{\romannumeral2} gives
a brief review of OMP and RIP, as well as certain necessary assumptions and notations. Section
\uppercase\expandafter{\romannumeral3} presents the main theoretical results on the completely
perturbed scenarios. Several extensions are also presented with respect to special signals. Section
\uppercase\expandafter{\romannumeral4} provides the proofs of the theorems. Section
\uppercase\expandafter{\romannumeral5} discusses some related works. The whole paper is concluded
in Section \uppercase\expandafter{\romannumeral6}. To make the paper more readable, some proofs are
relegated as an appendix in Section \uppercase\expandafter{\romannumeral7}.

\section{Background}

\subsection{Orthogonal Matching Pursuit (OMP)}

The key idea of OMP lies in the attempt to reconstruct the support set $\Lambda$  of $\bm x$
iteratively by starting with $\Lambda=\emptyset$. In the $l$th iteration, the inner products
between the columns of $\bm \Phi$ and the residual $\bm r^{l-1}$ are calculated, and the index of
the largest absolute value of inner products is added to $\Lambda$. Here, the residual $\bm
r^{l-1}$ from the former iteration represents the component of the measurement vector $\bm y$ that
cannot be spanned by the columns of $\bm \Phi$ indexed by $\Lambda$. In this way, the columns of
$\bm \Phi$ which are ``the most relative" to $\bm y$ are iteratively chosen. The OMP algorithm is
described in Table~\ref{ompalgorithm}. It is necessary to point out that the version of OMP in this
paper does not require $k$, which appears throughout the paper, to be an input. In fact, we are
just concerned with the performance of OMP at the $k$th iteration.

\begin{table}[t]
\renewcommand{\arraystretch}{1.4}
\renewcommand{\arraystretch}{1.2}
\caption{The OMP Algorithm}
\begin{center}
\begin{tabular}{l}
\hline {\bf Input:}\label{ompalgorithm}
\hspace{0.5em} $\bm y$, $\bm \Phi$; \\
{\bf Initialization:} \hspace{0.5em}
$\bm r^0=\bm y$, $\Lambda^0=\emptyset$, $l=0$;\\
{\bf Repeat}\\
\hspace{3.5em}$l=l+1$;\\
\hspace{1.5em}match step:\\
\hspace{3.5em}$\bm h^l=\bm \Phi^\textrm{T} \bm r^{l-1}$;\\
\hspace{1.5em}identify step:\\
\hspace{3.5em}$\Lambda^l=\Lambda^{l-1} \cup \{\textrm{arg}\,\max_j |\bm h^l(j)|\}$;\\
\hspace{1.5em}update step:\\
\hspace{3.5em}$\bm x^l=\textrm{arg}\,\min_{\bm z: \textrm{supp}(\bm
z)\subseteq\Lambda^l}\|\bm y- \bm \Phi \bm z\|_2$;\\
\hspace{3.5em}$\bm r^l=\bm y-\bm \Phi \bm x^l$;\\
{\bf Until} \hspace{0.5em} stop criterion satisfied;\\
{\bf Output:}
\hspace{0.5em}$\bm x^k$.\\
\hline
\end{tabular}
\end{center}
\end{table}

In fact, OMP can be well expressed using $\bm y$, $\bm \Phi$, $\Lambda^l$, Moore-Penrose
pseudoinverse, and orthogonal projection operator. A detailed analysis has been given in
\cite{Mark}. To introduce the case of noise and pave way for the proof of main results, a brief
review of them is given as follows.

Let $\bm u|_\Lambda$ denote the $|\Lambda|\times 1$ vector containing the entries of $\bm u$
indexed by $\Lambda$. Define $\bm u(j)$ as the $j$th entry of vector $\bm u$. Let $\bm
\Phi_\Lambda$ denote the $m \times |\Lambda|$ matrix obtained by selecting the columns of sensing
matrix $\bm \Phi$ indexed by $\Lambda$. If $\bm \Phi_\Lambda$ has full column rank, then $\bm
\Phi_\Lambda^\dag=(\bm \Phi_\Lambda^\textrm{T} \red{\bm \Phi_\Lambda})^{-1} \bm \Phi_\Lambda^\textrm{T}$
is the Moore-Penrose pseudoinverse of $\bm \Phi_\Lambda$. Let $\bm P_\Lambda = \bm \Phi_\Lambda \bm
\Phi_\Lambda^\dag$ and $\bm P_\Lambda^\bot = \bm I-\bm P_\Lambda$ denote the orthogonal projection
operator onto the column space of $\bm \Phi_\Lambda $ and its orthogonal complement, respectively.
Define $\bm A_\Lambda = \bm P_\Lambda^\bot \bm \Phi$ and $\bm A_\Lambda=\bm \Phi$ when
$\Lambda=\emptyset$, then $\bm A_\Lambda$ has the same size as $\bm \Phi$. From the theory of
linear algebra, any orthogonal projection operator $\bm P$ obeys $\bm P=\bm P^\textrm{T}=\bm P^2$
and the columns of $\bm A_\Lambda$ indexed by $\Lambda$ are zeros.

In the $l$th iteration, we begin with the estimation $\Lambda^{l-1}$ from the previous iteration.
The discussion below demonstrates the generation of $\Lambda^l$.

In the update step of the previous iteration, which is actually solving a least square problem, one
has
\begin{align}
\bm r^{l-1} &= \bm y-\bm \Phi \bm x^{l-1} =\bm y - \bm \Phi_{\Lambda^{l-1}} \bm
\Phi_{\Lambda^{l-1}}^\dag \bm y=\bm P_{\Lambda^{l-1}}^\bot \bm y.\label{dj14}
\end{align}
In the matching step, one has
\begin{align}\label{dj28}
\bm h^l &= \bm \Phi^\textrm{T} \bm r^{l-1} =\bm \Phi^\textrm{T} (\bm
P_{\Lambda^{l-1}}^\bot)^\textrm{T} \bm P_{\Lambda^{l-1}}^\bot \bm y = \bm
A_{\Lambda^{l-1}}^\textrm{T} \bm r^{l-1}.
\end{align}

From (\ref{dj14}), (\ref{dj28}), and the fact that the columns of $\bm A_\Lambda$ indexed by
$\Lambda$ are zeros, it can be derived that
\begin{equation}\label{dj27}
{\bm h}^l(j)=0,\ \ \ \forall j \in\Lambda^{l-1}.
\end{equation}
Therefore $\textrm{arg}\,\max_j |\bm h^l(j)|\notin \Lambda^{l-1}$, $|\Lambda^{l}|=l$.

It is important to notice that the above property still holds when $\bm y$ and $\bm \Phi$ are
replaced by the contaminated $\tilde{\bm y}$ and $\tilde{\bm \Phi}$. To see this, it is calculated
that
\begin{align}
{\bm r}^{l-1} &= \tilde{\bm y}-\tilde{\bm \Phi} {\bm x}^{l-1} =\tilde{\bm y} - \tilde{\bm
\Phi}_{\Lambda^{l-1}} \tilde{\bm \Phi}_{\Lambda^{l-1}}^\dag \tilde{\bm y}=\tilde{\bm
P}_{\Lambda^{l-1}}^\bot \tilde{\bm y},\label{r16}\\
{\bm h}^l &= \tilde{\bm \Phi}^\textrm{T} {\bm r}^{l-1} = \tilde{\bm \Phi}^\textrm{T} (\tilde{\bm
P}_{\Lambda^{l-1}}^\bot)^\textrm{T} \tilde{\bm P}_{\Lambda^{l-1}}^\bot \tilde{\bm y} = \tilde{\bm
A}_{\Lambda^{l-1}}^\textrm{T} {\bm r}^{l-1},\label{r17}
\end{align}
where $\tilde{\bm P}_{\Lambda}^\bot$ and $\tilde{\bm A}_{\Lambda}$ are defined by the perturbed
sensing matrix $\tilde{\bm \Phi}$. Due to the fact that the columns of $\tilde{\bm
A}_{\Lambda^{l-1}}$ indexed by $\Lambda^{l-1}$ still equal zeros, (\ref{dj27}) holds in the
completely perturbed scenario.

\subsection{The Restricted Isometry Property (RIP)}

For each integer $k=1,2,\ldots,n$, the RIP for any matrix $\bm A\in \mathbb{C}^{m\times n}$ defines
the restricted isometry constant (RIC) $\delta_k$ as the smallest nonnegative number such that
\begin{equation}
(1-\delta_k)\|\bm x\|_2^2 \leq \|\bm A \bm x\|_2^2 \leq (1+\delta_k) \|\bm x\|_2^2
\end{equation}
holds for any $k$-sparse vector $\bm x$ \cite{Tao}. It is easy to check that if $\bm A$ satisfies
the RIP of order $k_1$ and $k_2$ with isometry constants $\delta_{k_1}$ and $\delta_{k_2}$,
respectively, and $k_1\leq k_2$, then one has $\delta_{k_1} \leq \delta_{k_2}$.

Since the introduction of the RIP, it has been widely used as a tool to guarantee successful sparse
recovery for various algorithms. For example, for the $(N_0)$ process, the RIP of order $2k$ with
$\delta_{2k}<0.03/\sqrt{\log k}$ guarantees exact recovery for any $k$-sparse signal via ROMP
\cite{Deanna}; the RIP of order $3k$ with $\delta_{3k}<0.165$ permits SP  to
exactly recover any $k$-sparse signal \cite{Wei}.

However, analyzing the performance of OMP with RIP was relatively elusive before Davenport and
Wakin's work in \cite{Mark}. They demonstrated that RIP can be used for a very straightforward
analysis of OMP in $(N_0)$ process. It is shown that if $\bm y=\bm \Phi \bm x$ and $\bm x$ is a
$k$-sparse signal, then $\delta_{k+1}<\red{\frac{1}{3\sqrt{k}}}$ is sufficient for exact recovery of OMP
\cite{Mark} (Th.3.1). Later, Liu and Temlyakov relaxed the bound to $\red{\frac1{(1+\sqrt{2})\sqrt{k}}}$
\cite{Entao} (Th.5.2). Huang and Zhu further improved the bound to $\red{\frac1{1+2\sqrt{k}}}$, and they also
discussed the performance for the $(N_1)$ process \cite{review2}. In \cite{remark}, it has been
proved that $\red{\frac1{1+\sqrt{k}}}$ is sufficient for $(N_0)$ process, and for any given $k>1$, there
exists a sensing matrix with $\delta_{k+1}=1/\sqrt{k}$ and a $k$-sparse signal that exact recovery
via OMP is not guaranteed. Therefore, if one uses the RIP of order $k+1$ as a sufficient condition
for exact recovery of a sparse signal via OMP, little improvement is possible. In terms of the
number of measurements, for Gaussian or Bernoulli matrices it was demonstrated in \cite{Mark} that
$\delta_{k+1}<\red{\frac1{1+\sqrt{k}}}$ requires $\red{O(k^2 \log\frac{n}{k})}$ measurements, and the number is roughly
the same as what is required by coherence-based analysis in \cite{Tropp}.

\subsection{Assumptions and Notations}

A vector $\bm x\in\mathbb{C}^n$ is $k$-sparse if it contains no more than $k$ nonzero entries.
Throughout this paper, however, the signal to be recovered is not limited to a sparse one. For a
non-sparse signal $\bm x$, define $\bm x^{(1)}\in\mathbb{C}^n$ as the $k$-sparse signal that
contains the $k$ largest magnitude entries of $\bm x$ (i.e. the best $k$-term approximation of $\bm
x$), and define $\bm x^{(2)}=\bm x-\bm x^{(1)}$. In order to delineate the compressibility of a
general signal $\bm x$, define
\begin{equation*}
\beta=\frac{\|\bm x^{(2)}\|_2}{\|\bm x^{(1)}\|_2},\ \ \gamma=\frac{\|\bm x^{(2)}\|_1}{\sqrt{k}\|\bm
x^{(1)}\|_2}.
\end{equation*}
In this paper, $\bm x$ is assumed to be almost sparse (i.e. $\beta$ and $\gamma$ are far less than
$1$). When $\bm x^{(2)}=\bm 0$, one has $\beta=\gamma=0$, and $\bm x$ is reduced to a sparse
signal.

The notation of strong-decaying sparse signals is introduced by Davenport and Wakin in \cite{Mark}.
In our work, such concept is extended to general signals termed strong-decaying signals. Let
$\left\{\bm x(m_j)\right\}_{1\leq j\leq n}$ denote the entries of $\bm x$ rearranged in descending
order by magnitude. $\bm x$ is called an $\alpha$-strong-decaying signal if for all
$j\in\{1,2,\ldots n-1\}$ and $\bm x(m_{j+1})\neq 0$, $\quad |\bm x(m_j)|/|\bm x(m_{j+1})|\geq
\alpha$, where $\alpha >1$ is a constant.

When $(N_2)$ or $(N_2')$ process is concerned, it is necessary to consider the nature of $\bm b$
and $\bm E$, and how they influence the process of OMP. This leads to the following definitions of
relative bounds, which were introduced by Herman and Strohmer in \cite{Matthew}.

The symbols $\|\cdot\|_2$ and $\|\cdot\|_2^{(k)}$ denote the spectral norm of a matrix and the
largest spectral norm taken over all $k$-column submatrices, respectively. The noise $\bm
b$ and the perturbation $\bm E$ can be quantified as
\begin{equation}\label{dj16}
\frac{\|\bm b\|_2}{\|\bm \Phi \bm x\|_2}\leq \varepsilon_b,\ \ \frac{\|\bm E\|_2^{(k)}}{\|\bm
\Phi\|_2^{(k)}}\leq \varepsilon,
\end{equation}
where $\|\bm \Phi \bm x\|_2$, $\|\bm \Phi\|_2$, and $\|\bm \Phi\|_2^{(k)}$ are nonzero. These
relative upper bounds provide an access to analyze the influence of $\bm b$ and $\bm E$, even
though the exact forms of them are unknown. Throughout this paper, it is appropriate to assume that
$\varepsilon_b$ and $\varepsilon $ are far less than $1$.

\section{Contributions}

In this section, a completely perturbed scenario in the form of (\ref{dj13}) is considered and the
performance of OMP in $(N_2)$ process is studied. Theorem 1 presents the RIP-based conditions under
which the support set of the best $k$-term approximation of $\bm x$ can be exactly recovered. In
Theorem 2, we construct a sensing matrix and perturbations with which an almost sparse signal
cannot be recovered. The RIC of the matrix is slightly bigger than that in the conditions of
Theorem 1, which indicates that the sufficient conditions in Theorem 1 are rather tight. Several
extensions with respect to special signals such as strong-decaying signals are put forward in
Theorem 3 and 4. In Theorem 5, perturbations in the form of (\ref{r28}) is considered and the
performance of OMP in $(N_2')$ process is studied. The following theorems and remarks summarize the
main results.

\textbf{Theorem 1:} Suppose that the inputs $\bm y$ and $\bm \Phi$ of OMP are contaminated by
perturbations in the form of (\ref{dj13}), and that the original signal $\bm x$ is almost sparse.
Define the relative perturbations $\varepsilon_b$ and $\varepsilon$ as in (\ref{dj16}). Let $t_0 =
\min_{j\in \textrm{supp}(\bm x^{(1)})} \,|\bm x(j)| $, and
\begin{align}\label{dj33}
\varepsilon_h =\frac{1.23}{1-\varepsilon}(\varepsilon+\varepsilon_b
+(1+\varepsilon_b)(\beta+\gamma))\|\bm x^{(1)}\|_2.
\end{align}
If $\tilde{\bm \Phi}$ satisfies the RIP of order $k+1$ with isometry constant
\begin{equation}\label{dj17}
\tilde{\delta}_{k+1} < Q(k,\varepsilon_h/t_0),
\end{equation}
then OMP will recover the support set of $\bm x^{(1)}$ exactly from $\tilde{\bm y}$ and $\tilde{\bm
\Phi}$ in $k$ iterations, and the error between $\bm x^{(1)}$ and the recovered $k$-sparse signal
$\hat{\bm x}$ can be bounded as
\begin{equation}\label{r27}
\|\hat{\bm x}-\bm x^{(1)}\|_2 \leq \frac{\varepsilon_h}{\sqrt{1-\tilde{\delta}_k}}.
\end{equation}
In (\ref{dj17}) the function $Q(\cdot,\cdot)$ is defined as
\begin{equation}\label{dj61}
Q(u,v)=\frac{1}{\sqrt{u}+1}-\frac{3}{\sqrt{u}+1}v.
\end{equation}

\begin{proof}
The proof consists of three parts. The former two parts prove that
\begin{equation}\label{r1}
\tilde{\delta}_{k+1}<\frac{1}{\sqrt{k}+1}-\frac{3}{\sqrt{k}+1}\frac{\|\bm e\|_2}{t_0},
\end{equation}
where $\bm e=\tilde{\bm \Phi} \bm x^{(2)}-\bm E\bm x+\bm b$, is a sufficient condition for the
support recovery. The last part then gives an upper bound of $\|\bm e\|_2$, i.e. $\varepsilon_h$.

The detailed proof is postponed to Section \uppercase\expandafter{\romannumeral4}-B.
\end{proof}

\textit{Remark 1:} Theorem 1 reveals that if the RIC of the available sensing matrix $\tilde{\bm
\Phi}$ is known to be under a threshold, it is guaranteed that the support set of the best $k$-term approximation of a
signal can be recovered.

It is of great significance to properly interpret $\varepsilon_h/t_0$ in (\ref{dj17}). On one hand,
the effects of $\bm b$ and $\bm E$ are reflected in terms of the worst-case relative perturbation
$\varepsilon_b$ and $\varepsilon$, respectively. Therefore $\varepsilon_h$ represents a worst-case
effect from perturbed $\tilde{\bm y}$ and $\tilde{\bm \Phi}$. If more information on $\bm b$ and
$\bm E$ is known, it may be possible to estimate a smaller $\varepsilon_h$. On the other hand,
$t_0$ is the smallest magnitude of nonzero entries in $\bm x^{(1)}$ and represents the capability
of a sparse signal to be recovered against perturbations. Therefore, $t_0/\varepsilon_h$ has a
natural interpretation as a lower bound on the minimum component SNR. One can see that the bound on
$\tilde{\delta}_{k+1}$ increases as $t_0/e_h$ increases.

\textit{Remark 2:} Considering $(N_0)$ process, Theorem 1 generalizes the results in
\cite{Mark,Entao,review2,remark}. If vector $\bm y$ and matrix $\bm \Phi$ are unperturbed, and $\bm
x$ is $k$-sparse, then (\ref{dj17}) reduces to
\begin{equation}\label{dj23}
\delta_{k+1}<\frac{1}{\sqrt{k}+1},
\end{equation}
which is exactly the result in \cite{remark}.

\textit{Remark 3:} It needs to be pointed out that in order to be well defined,
$Q(k,\varepsilon_h/t_0)$ should be greater than zero. Thus one gets
\begin{equation}\label{dj18}
t_0>3\varepsilon_h.
\end{equation}
It means that for the best $k$-term approximation of an almost sparse signal, the lower bound on
the minimum component SNR should be large enough, so that its support can be extracted despite
various perturbations.

When $\bm x$ is $k$-sparse and only the measurement vector $\bm y$ is perturbed, two corollaries
can be derived from Theorem 1.

\textbf{Corollary 1:} Suppose that $\bm E=\bm 0$ in (\ref{dj13}) and that the original signal $\bm
x$ is $k$-sparse. Let
\begin{align*}
\varepsilon_h =1.23\varepsilon_b\|\bm x\|_2.
\end{align*}
If $\bm \Phi$ satisfies the RIP of order $k+1$ with isometry constant
\begin{equation}
\delta_{k+1} < Q(k,\varepsilon_h/t_0),
\end{equation}
then OMP will recover the support set of $\bm x$ exactly from $\tilde{\bm y}$ and $\bm \Phi$ in $k$
iterations, and the error between $\bm x$ and the recovered $k$-sparse signal $\hat{\bm x}$ can be
bounded as
\begin{equation}
\|\hat{\bm x}-\bm x\|_2 \leq \frac{\varepsilon_h}{\sqrt{1-\delta_k}}.
\end{equation}

\textbf{Corollary $\bm 1'$:} Suppose that $\bm E=\bm 0$ in (\ref{dj13}) and that the original
signal $\bm x$ is $k$-sparse. If $\bm \Phi$ satisfies the RIP of order $k+1$ with isometry constant
\begin{equation}\label{r24}
\delta_{k+1} < \frac{1-\tau}{\sqrt{k}+1},
\end{equation}
where $\tau\in (0,1)$ is a constant, and
\begin{equation}\label{r25}
\|\bm b\|_2 \leq \tau t_0 /3,
\end{equation}
then OMP will recover the support set of $\bm x$ exactly from $\tilde{\bm y}$ and $\bm \Phi$ in $k$
iterations, and the error between $\bm x$ and the recovered $k$-sparse signal $\hat{\bm x}$ can be
bounded as
\begin{equation}
\|\hat{\bm x}-\bm x\|_2 \leq \frac{\|\bm b\|_2}{\sqrt{1-\delta_k}}.
\end{equation}

\textit{Remark 4:} Both Corollary 1 and Corollary $1'$ concern the conditions for exact recovery of
${\rm supp}(\bm x)$ under measurement noise, but they are obtained from different point of views.
In Corollary $1'$, the bound of $\delta_{k+1}$ is unrelated with the noise for a given $\tau$,
while the $\ell_2$ norm of measurement noise should be under a threshold. A comparison of Corollary
1 with a similar conclusion \cite{BPN1} (Th.5.1), and a comparison of Corollary $1'$ with
conclusion \cite{review2} (Th.2) will be given in Section \uppercase\expandafter{\romannumeral5}.

When neither the measurement vector nor the sensing matrix is perturbed, the following corollary
gives sufficient conditions under which the support of the best $k$-term approximation of an almost
sparse signal can be exactly recovered. A similar conclusion in \cite{error} (Th.3.1) will be
compared with Corollary 2 in Section \uppercase\expandafter{\romannumeral5}.

\textbf{Corollary 2:} Suppose that $\bm b=\bm 0$, $\bm E=\bm 0$ in (\ref{dj13}), and that the
original signal $\bm x$ is almost sparse. Let
\begin{align}\label{r33}
\varepsilon_h =1.23(\beta+\gamma)\|\bm x^{(1)}\|_2.
\end{align}
If $\bm \Phi$ satisfies the RIP of order $k+1$ with isometry constant
\begin{equation}
\delta_{k+1} < Q(k,\varepsilon_h/t_0),
\end{equation}
then OMP will recover the support set of $\bm x^{(1)}$ exactly from $\bm y$ and $\bm \Phi$ in $k$
iterations, and the error between $\bm x^{(1)}$ and the recovered $k$-sparse signal $\hat{\bm x}$
can be bounded as
\begin{equation}
\|\hat{\bm x}-\bm x^{(1)}\|_2 \leq \frac{\varepsilon_h}{\sqrt{1-\delta_k}}.
\end{equation}

Inspired by the work \cite{remark}, the following theorem reveals how tight the RIP-based
conditions in Theorem 1 are.

\textbf{Theorem 2:} Consider the completely perturbed scenario (\ref{dj13}). For any given positive
integer $k\ge 2$, $\eta>1$, constants $t_0>0$ and $0<\xi<t_0$, there exist an almost sparse signal
$\bm x\in\mathbb{C}^{k+1}$, a sensing matrix $\bm \Phi\in\mathbb{C}^{(k+1)\times(k+1)}$,
perturbations $\bm E$ and $\bm b$ such that the smallest nonzero magnitude of $k$-sparse $\bm
x^{(1)}$ is $t_0$,
$$\xi=\|\bm e\|_2=\|\tilde{\bm \Phi} \bm x^{(2)}-\bm E\bm x+\bm b\|_2,$$
and the perturbed sensing matrix $\tilde{\bm \Phi}$ satisfies the RIP of order $k+1$ with isometry
constant
\begin{equation}\label{knex3}
\tilde \delta_{k+1} \le \frac{\eta}{\sqrt{k}}-\frac{\sqrt{k-1}}{k}\frac{\xi}{t_0}.
\end{equation}
Furthermore, OMP fails to recover the support set of $\bm x^{(1)}$ from $\tilde{\bm y}$ and
$\tilde{\bm \Phi}$ in $k$ iterations.

\begin{proof}
The proof is postponed to Section \uppercase\expandafter{\romannumeral4}-C.
\end{proof}

Compared to [27, Th.3.2], Theorem 2 takes general perturbations as well as
non-sparseness of $\bm x$ into consideration. Setting $\xi=0$ and $\eta\rightarrow 1$ in Theorem 2,
it reduces to the result in \red{\cite{remark}.}

\textit{Remark 5:} It will be shown that the bound (\ref{dj17}) is rather tight and little
improvement can be made on it. In the proof of Theorem 1, we first prove that (\ref{r1}) is a
sufficient condition for support recovery of ${\bm x}^{(1)}$, then estimate $\|\bm e\|_2$ by $\|\bm
e\|_2\le \varepsilon_h$. Comparing (\ref{knex3}) with (\ref{r1}), these two bounds are both linear
decreasing function of $\|\bm e\|_2/t_0$, and as $k$ tends to infinity, the ratio of their
$y$-intercepts approaches 1 as $\eta\rightarrow 1$, and the ratio of their slopes approaches 3. As
for the upper bound of $\|\bm e\|_2$, Proposition 3.5 in \cite{Needell} and triangle inequality are
used. Because the equality of the Proposition 3.5 in \cite{Needell} is difficult to be achieved, we
assume that $\bm x$ is $k$-sparse for the sake of briefness. In fact,
$$
\|\bm e\|_2=\frac{\|\tilde{\bm \Phi}\|_2^{(k)}}{1-\varepsilon}(\varepsilon+\varepsilon_b)\|\bm
x\|_2
$$
can be satisfied. First, let $\bm E=-\varepsilon \bm \Phi$, and choose a $k$-sparse signal $\bm x$
that satisfies $\|\tilde{\bm \Phi} \bm x\|_2=\|\tilde{\bm \Phi}\|_2^{(k)}\|\bm x\|_2$. Let $\bm
b=\varepsilon_b \bm \Phi\bm x$. Then it holds that
\begin{align*}
\|\bm e\|_2=\|-\bm E\bm x+\bm b\|_2=\|(\varepsilon+\varepsilon_b)\bm \Phi \bm x\|_2
=(\varepsilon+\varepsilon_b)\frac{\|\bm \tilde{\bm \Phi} \bm x\|_2}{1-\varepsilon}
=\frac{\|\tilde{\bm \Phi}\|_2^{(k)}}{1-\varepsilon}(\varepsilon+\varepsilon_b)\|\bm x\|_2.
\end{align*}
\red{Due} to the above two reasons, we show that the bound (\ref{dj17}) in Theorem 1 is rather tight.

For $\alpha$-strong-decaying signals, the requirement of isometry constant $\tilde{\delta}_{k+1}$
can be relaxed, and the locations can even be picked up in the order of their entries' magnitude as
long as the decaying constant $\alpha$ is large enough. This is what the following two theorems
reveal.

\textbf{Theorem 3:} Suppose that the inputs $\bm y$ and $\bm \Phi$ of OMP are
contaminated by perturbations as in (\ref{dj13}), and that the original signal $\bm x$ is
$\alpha$-strong-decaying. Let
\begin{align*}
\varepsilon_h =\frac{1.23}{1-\varepsilon}(\varepsilon+\varepsilon_b
+(1+\varepsilon_b)C\alpha^{-k})\|\bm x^{(1)}\|_2
\end{align*}
and $k^*=\red{\frac{(\sum_{i=0}^{k-1}\alpha^i)^2}{\sum_{i=0}^{k-1}\alpha^{2i}}}$, where $C$ is a constant
depending only on $\alpha$. If $\tilde{\bm \Phi}$ satisfies the RIP of order $k+1$ with isometry
constant
\begin{equation}\label{r12}
\tilde{\delta}_{k+1} < Q(k^*,\varepsilon_h/t_0),
\end{equation}
then OMP will recover the support set of $\bm x^{(1)}$ exactly from $\tilde{\bm y}$ and $\tilde{\bm
\Phi}$ in $k$ iterations, and the error between $\bm x^{(1)}$ and the recovered $k$-sparse signal
$\hat{\bm x}$ can be bounded as
\begin{equation}
\|\hat{\bm x}-\bm x^{(1)}\|_2 \leq \frac{\varepsilon_h}{\sqrt{1-\tilde{\delta_k}}}.
\end{equation}

\begin{proof}
The proof is postponed to Section \uppercase\expandafter{\romannumeral4}-D.
\end{proof}

\textit{Remark 6:} Theorem 3 reveals that the recovery of a strong-decaying signal needs relaxed
requirement of $\tilde{\delta}_{k+1}$, and that the larger $\alpha$ is, the easier the requirement
of $\tilde{\delta}_{k+1}$ can be satisfied. To see this, notice that for $k>1$, Cauchy-Schwarz
inequality implies $k^*<k$, and thus
\begin{equation*}
Q(k,\varepsilon_h/t_0)< Q(k^*,\varepsilon_h/t_0).
\end{equation*}
Define $k^*=L(\alpha)=(\sum_{i=0}^{k-1}\alpha^i)^2/\sum_{i=0}^{k-1}\alpha^{2i}$. Because
$L(\alpha)$ is a decreasing function of $\alpha$, the larger $\alpha$ is, the smaller $k^*$ is, and
the easier the requirement of $\tilde{\delta}_{k+1}$ can be satisfied.

\textbf{Corollary 3:} Suppose that the measurement vector and the sensing matrix are unperturbed, and
that the original signal $\bm x$ is a $k$-sparse $\alpha$-strong-decaying one. If $\bm \Phi$
satisfies the RIP of order $k+1$ with isometry constant
\begin{equation}\label{r11}
\delta_{k+1} < \frac{1}{\sqrt{k^*}+1},
\end{equation}
then OMP will recover $\bm x$ exactly from $\bm y$ and $\bm \Phi$ in $k$ iterations.

\textit{Remark 7:} Because $\red{k^*< \frac{\alpha+1}{\alpha-1}}$, the sufficient condition (\ref{r11}) can
be replaced by
\begin{equation*}
\delta_{k+1} < \frac{1}{\displaystyle \sqrt{\red{\frac{\alpha+1}{\alpha-1}}}+1}.
\end{equation*}
Furthermore, if $\alpha$ is far greater than $1$, the requirement approximately reduces to
$\delta_{k+1} < \red{0.5}$\red{.}

\textbf{Theorem 4:} Suppose that the inputs $\bm y$ and $\bm \Phi$ of OMP are
contaminated by perturbations as in (\ref{dj13}), and that the original signal $\bm x$ is
$\alpha$-strong-decaying. Let
\begin{align*}
\varepsilon_h =\frac{1.23}{1-\varepsilon}(\varepsilon+\varepsilon_b
+(1+\varepsilon_b)C\alpha^{-k})\|\bm x^{(1)}\|_2,
\end{align*}
where $C$ is a constant depending only on $\alpha$. If $\tilde{\bm \Phi}$ satisfies the RIP of
order $k+1$ with isometry constant
\begin{equation*}
\tilde{\delta}_{k+1} < \frac{1}{3} - \frac{2\varepsilon_h}{3t_0},
\end{equation*}
and
\begin{equation}\label{dj24}
\alpha\geq \max\left\{G(\tilde{\delta}_{k+1}),1.2\right\}
\end{equation}
where
\begin{equation}\label{defineg}
G(u) =\frac{1+u}{1-\displaystyle  3u - \frac{2\varepsilon_h}{t_0}},
\end{equation}
then OMP will recover the support set of $\bm x^{(1)}$ exactly from $\tilde{\bm y}$ and $\tilde{\bm
\Phi}$ in $k$ iterations, and the recovery is in the order of the signal entries' magnitude.

\begin{proof}
The proof is postponed to Section \uppercase\expandafter{\romannumeral4}-E.
\end{proof}

Based on Theorem~4, the following corollary can be directly derived. A comparison between this
corollary and a similar result in \cite{Mark} (Th.4.1) will be given in Section
\uppercase\expandafter{\romannumeral5}.

\textbf{Corollary 4:} Suppose that $\bm y$ and $\bm \Phi$ are unperturbed, and that the original
signal $\bm x$ is a $k$-sparse $\alpha$-strong-decaying one. If $\bm \Phi$ satisfies the RIP of
order $k+1$ with isometry constant
\begin{equation*}
\delta_{k+1} < \frac{1}{3},
\end{equation*}
and
\begin{equation}\label{r29}
\alpha>\max \left\{ \frac{1+\delta_{k+1}}{1-3 \delta_{k+1}}, 1.2 \right\},
\end{equation}
then OMP will recover $\bm x$ exactly from $\bm y$ and $\bm \Phi$ in $k$ iterations, and the
recovery is in the order of the entries' magnitude.

At the end of the main contribution, perturbations in the form of (\ref{r28}) is considered.

\textbf{Theorem 5:} Suppose that the inputs $\bm y$ and $\bm \Phi$ of OMP  are contaminated by
perturbations as in (\ref{r28}), and that the original signal $\bm x$ is almost sparse. Define the
relative perturbations $\varepsilon$ as that in (\ref{dj16}), and $\varepsilon_b$ as:
$$\frac{\|\bm b\|_2}{\|\tilde{\bm \Phi}\bm x\|_2}\leq \varepsilon_b.$$
Let
\begin{align*}
\varepsilon_h =1.23(\varepsilon+\varepsilon_b+\varepsilon \varepsilon_b
+(1+\varepsilon_b)(1+\varepsilon)(\beta+\gamma))\|\bm x^{(1)}\|_2.
\end{align*}
If ${\bm \Phi}$ satisfies the RIP of order $k+1$ with isometry constant
\begin{equation*}
{\delta}_{k+1} < Q(k,\varepsilon_h/t_0),
\end{equation*}
then OMP will recover the support set of $\bm x^{(1)}$ exactly from $\tilde{\bm y}$ and $\bm \Phi$
in $k$ iterations, and the error between $\bm x^{(1)}$ and the recovered $k$-sparse signal
$\hat{\bm x}$ can be bounded as
\begin{equation*}
\|\hat{\bm x}-\bm x^{(1)}\|_2 \leq \frac{\varepsilon_h}{\sqrt{1-{\delta_k}}}.
\end{equation*}

\begin{proof}
The proof is postponed to Section \uppercase\expandafter{\romannumeral4}-F.
\end{proof}

\textit{Remark 8:} The definition of $\varepsilon_b$ in Theorem 5 is different from that in Theorem
1. This is due to the fact that $\varepsilon_b$ denotes the relative measurement noise added to the
output of the system, and the output in this scenario is $\tilde{\bm \Phi}\bm x$ other than $\bm
\Phi \bm x$. By comparison of Theorem 1 and 5, it can be seen that their difference comes from the
respective definition of $\varepsilon_b$. Based on the completely perturbed scenario (\ref{r28}),
several results similar to Theorem 2-4 can be derived, and their proofs are analogous, thus they
are not included for simplicity.

\red{\section{Proofs}}

\red{\subsection{Lemmas}}

Before the proofs of the main theorems, two helpful lemmas are given first. Their proofs are
postponed to Appendix.

\textbf{Lemma 1:} Let $\{x_i\}_{1\le i\le l}$ denote $l$ positive variables satisfying
$x_i/x_{i-1}\geq \alpha$ for all $1<i\le l$, where $\alpha>1$ is a constant. Then the function
$$
f(x_1,x_2,\cdots,x_l)=\red{\frac{\sum\limits_{i=1}^l x_i^2}{\left(\sum\limits_{i=1}^l x_i\right)^{2}}}
$$
achieves its minimum value
$\red{\frac{\sum_{i=0}^{l-1}\alpha^{2i}}{\left(\sum_{i=0}^{l-1}\alpha^i\right)^{2}}}$ when $x_i/x_{i-1}=\alpha$,
$i=2,\cdots,l$.

\textbf{Lemma 2:} Suppose that the inputs $\bm y$ and $\bm \Phi$ of OMP  are contaminated by
perturbations as in (\ref{dj13}), and that the original signal $\bm x$ is an
$\alpha$-strong-decaying one. Let
\begin{align*}
\varepsilon_h =\frac{1.23}{1-\varepsilon}(\varepsilon+\varepsilon_b
+(1+\varepsilon_b)C\alpha^{-k})\|\bm x^{(1)}\|_2.
\end{align*}
For the $l$th iteration, define $\bm x^* \in  \mathbb{C}^n $ as the signal that contains the
entries of $\bm x$ indexed by $\textrm{supp}(\bm x^{(1)}) \setminus \Lambda^{l-1}$ with the rest
setting to zeros. If $\bm \tilde{\bm \Phi}$ satisfies the RIP of order $k+1$ with isometry constant
$\tilde{\delta}_{k+1}$, one has
\begin{equation}
|{\bm h}^l(j) - \bm x^*(j)| \leq \frac{\tilde{\delta}_{k+1}\|\bm
x^*\|_2+\varepsilon_h}{1-\tilde{\delta}_{k+1}}
\end{equation}
for all $j \notin \Lambda^{l-1} $.

\red{\subsection{Proof of Theorem 1}}

\begin{proof}
First of all, it will be proved that OMP exactly recovers the support set of $\bm x^{(1)}$ in $k$
iterations. This proof consists of three parts. First, we prove that (\ref{r1})
implies
\begin{equation}\label{knex1}
\tilde{\delta}_{k+1}<\frac{1}{\sqrt{k'}+1}-\frac{\sqrt{k'}+2}{\sqrt{k'}+1}\frac{\|\bm
e\|_2}{\sqrt{k'}t_0}
\end{equation}
for all $1\le k'\le k$. Second, define $\bm e=\tilde{\bm \Phi} \bm x^{(2)}-\bm E\bm x+\bm b$. We
prove that (\ref{knex1}) is a sufficient condition for the support recovery in the $l$th iteration
with $k'=k-l+1$. At last, an upper bound of $\|\bm e\|_2$ is given.

First, define
\begin{align}
c_1 = \frac{2+\sqrt{k'}}{(1+\sqrt{k})\sqrt{k'}t_0}, \ \ c_2 =
\frac{2+\sqrt{k'}}{(1+\sqrt{k'})\sqrt{k'}t_0},
\end{align}
then it's easy to check that $c_1<c_2$. According to (\ref{r1}), it can be derived that
\begin{align}
\tilde{\delta}_{k+1}<\frac{1}{\sqrt{k}+1}-c_1\|\bm
e\|_2\le\frac{c_2}{c_1}\left(\frac{1}{\sqrt{k}+1}-c_1\|\bm e\|_2\right)
=\frac{1}{\sqrt{k'}+1}-\frac{\sqrt{k'}+2}{\sqrt{k'}+1}\frac{\|\bm e\|_2}{\sqrt{k'}t_0},
\end{align}
which implies (\ref{knex1}).

The proof of the second part works by induction. To begin with, consider the first iteration where
$\Lambda^0=\emptyset$. (\ref{dj13}) indicates that
\begin{align}\label{r18}
\tilde{\bm y}&=(\tilde{\bm \Phi}-\bm E)\bm x+\bm b=\tilde{\bm \Phi}\bm x^{(1)}+\bm e.
\end{align}
Then,
\begin{equation*}
{\bm h}^1=\tilde{\bm \Phi}^\textrm{T} \bm \tilde{\bm y}=\tilde{\bm \Phi}^\textrm{T}(\tilde{\bm
\Phi}\bm x^{(1)}+\bm e),
\end{equation*}
which can be rewritten as
\begin{equation*}
{\bm h}^1(i)= \langle\tilde{\bm \Phi}\bm e_i,\tilde{\bm \Phi}\bm x^{(1)}+\bm e\rangle,
\end{equation*}
where $\langle\cdot,\cdot\rangle$ denotes the inner product in Euclidean space and $\bm e_i$
denotes the $i$th natural basis. Define
\begin{equation*}
H=\max_{i\in \textrm{supp}(\bm x^{(1)})}|{\bm h}^1(i)|
\end{equation*}
and $U=\big|\langle\tilde{\bm \Phi} \bm x^{(1)},\tilde{\bm \Phi} \bm x^{(1)}+\bm e\rangle\big|$. On
one hand,
\begin{align}\label{r9}
U&=\left|\sum \bm x^{(1)}(i) {\bm h}^1(i)\right|\leq \|\bm x^{(1)}\|_1 H \leq \sqrt{k}\|\bm
x^{(1)}\|_2 H.
\end{align}
On the other hand,
\begin{align}
U&\geq \|\tilde{\bm \Phi} \bm x^{(1)}\|_2^2-\|\tilde{\bm \Phi} \bm x^{(1)}\|_2 \|\bm e\|_2\nonumber\\
&\geq (1-\tilde{\delta}_{k+1})\|\bm x^{(1)}\|_2^2 - \sqrt{1+\tilde{\delta}_{k+1}}\|\bm
x^{(1)}\|_2\|\bm e\|_2.
\end{align}
Thus one has
\begin{equation}\label{r2}
H \geq \frac{1}{\sqrt{k}}\left((1-\tilde{\delta}_{k+1})\|\bm x^{(1)}\|_2 -
\sqrt{1+\tilde{\delta}_{k+1}}\|\bm e\|_2\right).
\end{equation}

For $i \notin \textrm{supp}(\bm x^{(1)})$, Lemma 2.1 in \cite{Emmanuel} implies that
\begin{align}
|{\bm h}^1(i)|&=|\langle\tilde{\bm \Phi}\bm e_i,\tilde{\bm \Phi}\bm
x^{(1)}\rangle+\langle\tilde{\bm \Phi}\red{\bm e_i},\bm e\rangle|\nonumber\\
&\leq \tilde{\delta}_{k+1}\|\bm x^{(1)}\|_2 + \|\tilde{\bm \Phi}\bm e_i\|_2 \|\bm e\|_2\nonumber\\
&\leq \tilde{\delta}_{k+1}\|\bm x^{(1)}\|_2 + \sqrt{1+\tilde{\delta}_{k+1}} \|\bm e\|_2.\label{r4}
\end{align}
Because $\|\bm x^{(1)}\|_2 \ge \sqrt{k}t_0$, (\ref{knex1}) of $k'=k$ together with (\ref{r2}) and
(\ref{r4}) indicate that
\begin{equation*}
H>|{\bm h}^1(i)|,\ \ \forall i\notin \textrm{supp}(\bm x^{(1)}),
\end{equation*}
which guarantees the success of the first iteration.

Now consider the general induction step. In the $l$th iteration, suppose that all previous
iterations succeed, which means that $\Lambda^{l-1}$ is a subset of ${\rm supp}(\bm x^{(1)})$.
Define $\bm z^{l-1}=\bm x^{(1)}-\bm {\bm x}^{l-1}$, then ${\rm supp}(\bm z^{l-1})\subseteq {\rm
supp}(\bm x^{(1)})$. Because
\begin{align*}
{\bm h}^l=\tilde{\bm \Phi}^\textrm{T}(\tilde{\bm y}-\tilde{\bm \Phi}\bm x^{l-1}) =\tilde{\bm
\Phi}^\textrm{T}(\tilde{\bm \Phi}\bm x^{(1)}-\tilde{\bm \Phi}\bm x^{l-1}+\bm e),
\end{align*}
one has
\begin{align*}
{\bm h}^l(i)= \langle\tilde{\bm \Phi}\bm e_i,\tilde{\bm \Phi}\bm z^{l-1}+\bm e\rangle.
\end{align*}
Define
\begin{equation*}
H=\max_{i\in \textrm{supp}(\bm x^{(1)})}|{\bm h}^l(i)|
\end{equation*}
and $U=\big|\langle\tilde{\bm \Phi} \bm z^{l-1},\tilde{\bm \Phi} \bm z^{l-1}+\bm e\rangle\big|$.
According to (\ref{dj27}), it can be derived that
\begin{align}\label{r8}
U&=\left|\sum \bm z^{l-1}(i) {\bm h}^l(i)\right|\nonumber\\
&\leq \|\bm z^{l-1}|_{{\rm supp}(\bm
x^{(1)})\setminus \Lambda^{l-1}}\|_1 H \nonumber\\
&\leq \sqrt{k-l+1}\|\bm z^{l-1}\|_2 \red{H} \nonumber\\
&\red{=}\sqrt{k'}\|\bm z^{l-1}\|_2 H.
\end{align}
Following the steps in the proof for the first iteration, and noticing that $\|\bm z^{l-1}\|_2 \ge
\sqrt{k'}t_0$, it can be derived from (\ref{knex1}) that
\begin{align*}
H > |{\bm h}^l(i)|,\ \ \forall i \notin \textrm{supp}(\bm x^{(1)}).
\end{align*}
According to (\ref{dj27}), ${\bm h}^l(i)=0$ for $i \in \Lambda^{l-1}$, which guarantees the success
of the $l$th iteration. The proof of induction is completed.

Thirdly, an upper bound of $\|\bm e\|_2$ is given as follows. According to Proposition 3.5 in
\cite{Needell},
\begin{align*}
\|\tilde{\bm \Phi} \bm x^{(2)}\|_2 &\leq \sqrt{1+\tilde{\delta}_k}(\|\bm x^{(2)}\|_2+\frac{\|\bm
x^{(2)}\|_1}{\sqrt{k}})\\
&=\sqrt{1+\tilde{\delta}_k}(\beta+\gamma)\|\bm x^{(1)}\|_2,
\end{align*}
and
\begin{align*}
\|\bm E \bm x\|_2 &\leq \|\bm E \bm x^{(1)}\|_2+\|\bm E \bm x^{(2)}\|_2\\
&\leq\|\bm E\|_2^{(k)}(\|\bm x^{(1)}\|_2+\|\bm x^{(2)}\|_2+\frac{\|\bm x^{(2)}\|_1}{\sqrt{k}})\\
&\leq \frac{\varepsilon}{1-\varepsilon}\sqrt{1+\tilde{\delta_k}} (1+\beta+\gamma)\|\bm x^{(1)}\|_2.
\end{align*}
Therefore,
\begin{align}
\|\bm e\|_2 &\leq \|\tilde{\bm \Phi} \bm x^{(2)}\|_2+\|\bm E\bm x\|_2+\|\bm b\|_2\nonumber\\
&\leq \|\tilde{\bm \Phi} \bm x^{(2)}\|_2+\|\bm E\bm x\|_2+\varepsilon_b \|\bm \Phi \bm x\|_2\nonumber\\
&\leq (\|\tilde{\bm \Phi} \bm x^{(2)}\|_2+\|\bm E\bm x\|_2 )(1+\varepsilon_b) + \varepsilon_b
\|\tilde{\bm \Phi} \bm x^{(1)}\|_2\nonumber\\
&\leq \frac{\sqrt{1+\tilde{\delta_k}}}{1-\varepsilon}(\varepsilon+\varepsilon_b
+(1+\varepsilon_b)(\beta+\gamma))\|\bm x^{(1)}\|_2.
\end{align}
Noticing that $\tilde{\delta_k}\leq \red{\frac1{\sqrt{k}+1}}\leq 0.5$, one has $\|\bm e\|_2 \le
\varepsilon_h$. Therefore (\ref{dj17}) implies (\ref{r1}), which guarantees the exact recovery of
${\rm supp}(\bm x^{(1)})$.

To finish the proof, the recovery error is bounded as follows. Because $\Lambda={\rm supp}(\bm
x^{(1)})$ is exactly recovered, one has
\begin{equation}\label{r26}
\hat{\bm x}|_\Lambda = \tilde {\bm \Phi}_\Lambda^\dag \tilde{\bm y} = \tilde {\bm
\Phi}_\Lambda^\dag (\tilde{\bm \Phi}_\Lambda \bm x^{(1)}|_\Lambda+\bm e)=\bm x^{(1)}|_\Lambda +
\tilde {\bm \Phi}_\Lambda^\dag \bm e.
\end{equation}
Thus
\begin{align*}
\|\hat{\bm x}-\bm x^{(1)} \|_2 &\leq \|\tilde {\bm \Phi}_\Lambda^\dag\|_2 \|\bm e\|_2 \leq
\frac{\varepsilon_h}{\sqrt{1-\tilde{\delta}_k}}.
\end{align*}
\end{proof}

\red{\subsection{Proof of Theorem 2}}

\begin{proof}
First, we prove that there exist a $k$-sparse signal $\bm x^{(1)}$ with $t_0$ as its smallest
nonzero entries' magnitude, a vector $\bm e\in\mathbb{C}^{k+1}$ satisfying $\|\bm e\|_2=\xi$, and a
perturbed sensing matrix $\tilde{\bm \Phi}$ with
\begin{align}\label{knex4}
\tilde \delta_{k+1} \le \frac{\eta}{\sqrt{k}}-\frac{\sqrt{k-1}}{k}\frac{\xi}{t_0}
\end{align}
such that OMP fails to recover the support set of $\bm x^{(1)}$ from $\tilde{\bm \Phi}$ and
$\tilde{\bm y}=\tilde{\bm \Phi}\bm x^{(1)}+\bm e$ in $k$ iterations if $\eta>1$. Let
\begin{align}\label{knex5}
\tilde{\bm \Phi}=\left(\begin{array}{cc}
                  \bm I_{k\times k} & a\bm 1_{k\times 1}\\
                  \bm 0_{1\times k} & b
                  \end{array}\right),
\end{align}
where $a=\delta/\sqrt{k}$ and $b=\sqrt{1-\delta^2}$ are two constants with $\delta<1/\sqrt{k}$.
Since
\begin{align}
\tilde{\bm \Phi}^{\rm T}\tilde{\bm \Phi}=\left(\begin{array}{cc}
                  \bm I_{k\times k} & a\bm 1_{k\times 1}\\
                  a\bm 1_{1\times k} & a^2k+b^2
                  \end{array}\right),
\end{align}
it can be derived that the eigenvalues $\{\lambda_i\}_{i=1}^{k+1}$ of $\tilde{\bm \Phi}^{\rm
T}\tilde{\bm \Phi}$ are
\begin{align}
\lambda_i=1,\ 1\le i\le k-1,\ \lambda_k=1-\delta,\ \lambda_{k+1}=1+\delta.
\end{align}
Thus for $\tilde{\bm \Phi}$, its RIC of order $k+1$ satisfies
\begin{align}
\tilde \delta_{k+1}=\delta.
\end{align}

Let $\bm x^{(1)} = (t_0\bm 1_{1\times k},0)^{\rm T}$ and $\bm e = (\bm 0_{1\times k},\xi)^{\rm T}$,
then the perturbed measurement vector
\begin{align*}
\tilde{\bm y}=\tilde{\bm \Phi}\bm x^{(1)}+\bm e=(t_0\bm 1_{1\times k},\xi)^{\rm T}.
\end{align*}
Set
\begin{align*}
\delta=\frac{\eta\sqrt{k}-(\xi/t_0)\sqrt{k-\eta^2+(\xi/t_0)^2}}{k+(\xi/t_0)^2},
\end{align*}
then the matching vector $\bm h^1=(t_0\bm 1_{1\times k},\eta t_0)^{\rm T}$, which implies that OMP
fails in the first iteration if $\eta>1$. It is easy to check that
\begin{align*}
\delta\le \frac{\eta}{\sqrt{k}}-\frac{\sqrt{k-1}}{k}\frac{\xi}{t_0}.
\end{align*}

Second, let $\bm \Phi=\bm I_{(k+1)\times(k+1)}$, $\bm x^{(2)}=(\bm 0_{1\times k}, \xi/2)^{\rm T}$,
$\bm b=(\bm 0_{1\times k}, \xi/2)^{\rm T}$, then
\begin{align*}
\bm E=\left(\begin{array}{cc}
                  \bm 0_{k\times k} & a\bm 1_{k\times 1}\\
                  \bm 0_{1\times k} & b-1
                  \end{array}\right),
\end{align*}
and $\bm e=\tilde{\bm \Phi}\bm x^{(2)}-\bm E\bm x+\bm b$, which completes the proof of Theorem 2.
\end{proof}

\red{\subsection{Proof of Theorem 3}}

\begin{proof}
The proof of Theorem 3 is similar to that of Theorem 1. For the sake of briefness, some revisions
are made based on the proof of Theorem 1.

First, define
\begin{align*}
k'^{*}=\red{\frac{\left(\sum_{i=0}^{k'-1}\alpha^i\right)^2}{\sum_{i=0}^{k'-1}\alpha^{2i}}}.
\end{align*}
According to Lemma 1, for any $\alpha$-strong-decaying and $k'$-sparse signal $\bm u$, it holds
that $\|\bm u\|_1\leq \sqrt{k'^{*}}\|\bm u\|_2$. Therefore, (\ref{r9}) and (\ref{r8}) can be
replaced by
\begin{align*}
U\leq \|\bm x^{(1)}\|_1 H\leq \|\bm x^{(1)}\|_2 \sqrt{k'^*}H
\end{align*}
and
\begin{align*}
U&\leq \|\bm z^{l-1}|_{{\rm supp}(\bm x^{(1)})\setminus \Lambda^{l-1}}\|_1 H \\
&=\|\bm x^{(1)}|_{{\rm
supp}(\bm x^{(1)})\setminus \Lambda^{l-1}}\|_1 H\\
&\leq \sqrt{k'^{*}}\|\bm x^{(1)}|_{{\rm supp}(\bm x^{(1)})\setminus \Lambda^{l-1}}\|_2 H\\
&\leq
\sqrt{k'^{*}}\|\bm z^{l-1}\|_2 H,
\end{align*}
respectively. Further, since
\begin{align*}
\|\bm z^{l-1}\|_2 \ge (\sum_{i=0}^{k'-1}\alpha^{2i})^{1/2} t_0\ge \sqrt{k'^{*}}t_0,
\end{align*}
equation (\ref{knex1}) can be replaced by
\begin{align}\label{knex2}
\tilde{\delta}_{k+1}<\frac{1}{\sqrt{k'^{*}}+1}-\frac{\sqrt{k'^{*}}+2}{\sqrt{k'^{*}}+1}\frac{\|\bm
e\|_2}{\sqrt{k'^{*}}t_0}.
\end{align}
Since $1\le k'^{*}\le k^*$, (\ref{knex2}) can be inferred by
\begin{align}
\tilde{\delta}_{k+1}<\frac{1}{\sqrt{k^{*}}+1}-\frac{3}{\sqrt{k^{*}}+1}\frac{\|\bm e\|_2}{t_0}.
\end{align}

Second, an upper bound of $\beta+\gamma$ can be given in terms of $\alpha$ as follows
\begin{align*}
\beta+\gamma &\leq \frac{|\bm x(m_{k+1})|((\sum_{i=0}^\infty \alpha^{-2i})^{1/2}+\sum_{i=0}^\infty
\alpha^{-i}/\sqrt{k})}{|\bm x(m_{k+1})| \sqrt{\sum_{i=1}^k \alpha^{2i}}}\\
&=\frac{(1-\alpha^{-2})^{-1/2}+(\sqrt{k}(1-\alpha^{-1}))^{-1}}{\alpha(\alpha^{2k}-1)^{1/2}(\alpha^2-1)^{-1/2}}\\
&=(1+\frac{1}{\sqrt{k}} \sqrt{\frac{\alpha+1}{\alpha-1}}) (\alpha^{2k}-1)^{-1/2}\\
&\leq C \alpha^{-k},
\end{align*}
where $C$ is a constant only related to $\alpha$. Therefore,
\begin{align*}
\|\bm e\|_2&\leq \frac{\sqrt{1+\tilde{\delta_k}}}{1-\varepsilon}(\varepsilon+\varepsilon_b +(1+\varepsilon_b)(\beta+\gamma))\|\bm x^{(1)}\|_2\\
&\leq \frac{\sqrt{1+\tilde{\delta_k}}}{1-\varepsilon}(\varepsilon+\varepsilon_b +(1+\varepsilon_b)C \alpha^{-k})\|\bm x^{(1)}\|_2.
\end{align*}
Notice that $\tilde{\delta_k}\leq \frac1{\red{\sqrt{k^*}+1}}\leq 0.5$, therefore (\ref{r12}) guarantees exact
recovery of ${\rm supp}(\bm x^{(1)})$.
\end{proof}

\red{\subsection{Proof of Theorem 4}}

\begin{proof}
By induction it will be shown that (\ref{dj24}) guarantees the order of recovery. For the $l$th
iteration, suppose that all the locations recovered in the previous iterations are in order. Define
$\bm x^*$ as that in Lemma 2. It will be demonstrated that OMP will choose the largest entry of
$\bm x^*$ (i.e. $\bm x(m_l)$). According to Lemma 2,
\begin{equation}\label{r22}
|{\bm h}^{l}(j)-\bm x^*(j)| \leq \frac{\tilde{\delta}_{k+1}\|\bm x^*\|_2 +
\varepsilon_h}{1-\tilde{\delta}_{k+1}}.
\end{equation}
It can be calculated from $\alpha \geq 1.2$ that
$$\sqrt{\frac{1}{1-\alpha^{-2}}}<1+\frac{1}{\alpha}.$$
Thus,
\begin{align}\label{r23}
\|\bm x^*\|_2 < |\bm x(m_l)| (\sum_{i=0}^\infty \alpha^{-2i})^{1/2} < |\bm x(m_l)|
(1+\frac{1}{\alpha}).
\end{align}
Combining (\ref{r22}) and (\ref{r23}), one has
\begin{align*}
|{\bm h}^{l}(m_l)| >& |\bm x(m_l)| - \Delta, \\
|{\bm h}^{l}(m_{j})| <& |\bm x(m_{l+1})| + \Delta \leq |\bm x(m_l)|/\alpha + \Delta, 
\quad j\in \{l+1,l+2,\ldots,n\},
\end{align*}
where
$$
\Delta = \frac{1}{1-\tilde{\delta}_{k+1}} (\tilde{\delta}_{k+1}|\bm x(m_l)| (1 +
\frac{1}{\alpha})+ \varepsilon_h).
$$
It is easy to check that $|{\bm h}^{l}(m_l)|$ is greater than $|{\bm h}^{l}(m_{j})|$ for $j\in
\{l+1,l+2,\ldots,n\}$, if (\ref{dj24}) is satisfied.
\end{proof}

\red{\subsection{Proof of Theorem 5}}

\begin{proof}
For the sake of briefness, we only need to make some revisions based on the proof of Theorem 1.
Noticing that the input $\bm \Phi$ is unperturbed and $\tilde{\bm y}=\tilde{\bm \Phi}\bm x+\bm
b=\bm \Phi \bm x^{(1)}+\bm \Phi \bm x^{(2)}+ \bm E \bm x+\bm b$, $\tilde{\delta}_{k+1}$ and $\bm e$
in the proof of Theorem 1 need to be replaced by $\delta_{k+1}$ and $\bm \Phi \bm x^{(2)}+ \bm E
\bm x+\bm b$.

Define $\bm e = \bm \Phi \bm x^{(2)}+ \bm E \bm x+\bm b$. An upper bound of $\|\bm e\|_2$ is given
as follows. According to Proposition 3.5 in \cite{Needell},
\begin{align*}
\|\bm \Phi \bm x^{(2)}\|_2 &\leq \sqrt{1+\delta_k}(\|\bm x^{(2)}\|_2+\frac{\|\bm
x^{(2)}\|_1}{\sqrt{k}})\\
&=\sqrt{1+\delta_k}(\beta+\gamma)\|\bm x^{(1)}\|_2,
\end{align*}
and
\begin{align*}
\|\bm E \bm x\|_2 &\leq \|\bm E \bm x^{(1)}\|_2+\|\bm E \bm x^{(2)}\|_2\\
&\leq \|\bm E\|_2^{(k)} (\|\bm x^{(1)}\|_2+\|\bm x^{(2)}\|_2+\frac{\|\bm x^{(2)}\|_1}{\sqrt{k}})\\
&= \varepsilon \sqrt{1+\delta_k} (1+\beta+\gamma)\|\bm x^{(1)}\|_2.
\end{align*}
Therefore,
\begin{align}
\|\bm e\|_2 &\leq \|\bm \Phi \bm x^{(2)}\|_2+\|\bm E\bm x\|_2+\|\bm b\|_2\nonumber\\
&\leq \|\bm \Phi \bm x^{(2)}\|_2+\|\bm E\bm x\|_2+\varepsilon_b \|\tilde{\bm \Phi} \bm x\|_2 \nonumber\\
&\leq (\|\bm \Phi \bm x^{(2)}\|_2+\|\bm E\bm x\|_2 )(1+\varepsilon_b) + \varepsilon_b \|\bm \Phi
\bm x^{(1)}\|_2\nonumber\\
&\leq \varepsilon_h,\nonumber
\end{align}
and
\red{\begin{align*}
\|\hat{\bm x} - \bm x^{(1)} \|_2\leq \|{\bm \Phi}_\Lambda^\dag\|_2 \|\bm e\|_2\leq
\frac{\varepsilon_h}{\sqrt{1-{\delta}_k}}.
\end{align*}}
\end{proof}

\red{\section{Related Works}}

In this section, Corollary 1, Corollary $1'$, Corollary 2, and Corollary 4 are compared with four
related conclusions in previous works.

\red{\subsection{Corollary 1 and [28 Th.5.1]}}

[28 Th.5.1]: Suppose that $\bm E=\bm 0$ in (\ref{dj13}) and $\bm x$ is $k$-sparse.
Define the coherence parameter $\mu$ of $\bm \Phi$ as $\mu=\max_{1\leq i,j\leq n,i\neq j}|\bm
G(i,j)|$, where $\bm G=\bm \Phi^{\textrm{T}} \bm \Phi$. If $\bm \Phi$ satisfies
\begin{equation}\label{r3}
k \leq \frac{1+\mu}{2 \mu}-\frac{1}{\mu}\frac{\|\bm b\|_2}{t_0},
\end{equation}
then OMP will recover the support set of $\bm x$ exactly and the recovery error can be bounded as
\begin{equation}\label{r3r}
\|\hat{\bm x}-\bm x\|_2 \leq \frac{\|\bm b\|_2}{\sqrt{1-\mu(k-1)}}.
\end{equation}

If we do not approximate the upper bound of $\|\bm e\|_2$ in terms of $\varepsilon_h$ in the proof
of Theorem 1, Corollary 1 derived from Theorem 1 has a more relaxed expression:

\textbf{Corollary 1$^*$:} Suppose that $\bm E=\bm 0$ in (\ref{dj13}) and $\bm x$ is $k$-sparse.
If $\bm \Phi$ satisfies the RIP of order $k+1$ with isometry constant
\begin{equation}\label{r24}
\delta_{k+1} < Q(k,\|\bm e\|_2/t_0)=\frac{1-\frac{3\|\red{\bm b}\|_2}{t_0}}{\sqrt{k}+1},
\end{equation}
then OMP will recover the support set of $\bm x$ exactly and the error can be bounded as
\begin{equation}\label{r24r}
\|\hat{\bm x}-\bm x\|_2 \leq \frac{\|\bm b\|_2}{\sqrt{1-\delta_k}}.
\end{equation}

Although Theorem 5.1 in \cite{BPN1} is coherence-based while Corollary 1$^*$ is RIC-based, they
both provide conditions for successful support recovery under measurement noise, based on which the
recovery error is further estimated. The comparisons are conducted from two aspects.

First, consider the ratio of the upper bounds on the recovery error in (\ref{r24r}) and
(\ref{r3r}):
$$
r=\frac{\frac{\|\red{\bm b}\|_2}{\sqrt{1-\delta_k}}}{\frac{\|\red{\bm
b}\|_2}{\sqrt{1-\mu(k-1)}}}=\frac{\sqrt{1-\mu(k-1)}}{\sqrt{1-\delta_k}}.
$$
According to Proposition 4.1 in \cite{Guangwu}, $\delta_{k}\leq \mu(k-1)$, and thus $r\leq 1$. This
means that the error bound given by Corollary 1$^*$ is at least as good as that in  [28, Th.5.1].

Second, consider the sufficient conditions for successful support recovery of the two results.
Direct comparison between (\ref{r3}) and (\ref{r24}) is difficult since as far as we know, there is
no clear comparison between $\delta_{k+1}$ and $\sqrt{k}\mu$ for arbitrary sensing matrix. For
simplicity, consider the scenario that the sensing matrix is Gaussian, and $m$, $n$, and $k$
increase in a proportional manner, i.e. $m/n \rightarrow \omega$ and $k/m \rightarrow \rho$ as $m
\rightarrow \infty$, where $\omega,\rho \in [0,1]$ are two constants. Results in \cite{Tanner} show
that there exists a constant $\delta(\omega,\rho)$ such that $\delta_{k+1}\le\delta(\omega,\rho)$
with high probability. Another result in \cite{Joel} reveals that $\mu\ge\sqrt{cm^{-1}\ln n}$ holds
with high probability where $c$ is a constant. Thus
$\sqrt{k}\mu\ge\sqrt{c\rho\ln(\omega^{-1}\rho^{-1}k)}$ with high probability. Inequality (\ref{r3})
implies that
\begin{align}\label{knex11}
\frac{\|\bm b\|_2}{t_0}\le\frac{1+\mu}{2}-\mu k\le
1-\sqrt{c\rho}\sqrt{k\ln(\omega^{-1}\rho^{-1}k)},
\end{align}
and the following inequality
\begin{align}\label{knex12}
\frac{\|\bm b\|_2}{t_0}\le\frac{1-\delta(\omega,\rho)}{3}-\frac{\delta(\omega,\rho)}{3}\sqrt{k}
\end{align}
implies (\ref{r24}). Since the bound in (\ref{knex11}) decreases with a higher order than that in
(\ref{knex12}) as $k$ increases, the sufficient condition (\ref{r24}) is more relaxed in this
sense.

\red{\subsection{Corollary $1'$ and [25, Th.2]}}

In \cite{review2}, it is proved that for $(N_1)$ process, the support of a $k$-sparse signal $\bm
x$ can be recovered, provided that
$$\delta_{k+1}<\frac1{\red{1+(\sqrt{6}+2)\sqrt{k}}}$$
and $\|\bm b\|_2\leq \delta_k \sqrt{k}t_0$. By comparison, it is shown that Corollary $1'$ is at
least as good as this conclusion.

First, let $\tau$ satisfy $\frac1{\red{1+(\sqrt{6}+2)\sqrt{k}}}=\red{\frac{1-\tau}{\sqrt{k}+1}}$, i.e.
$$\tau =\frac{(\sqrt{6}+1)\sqrt{k}}{1+(\sqrt{6}+2)\sqrt{k}}.
$$
Consider the ratio of the required upper bound of $\|\bm b\|_2$ in the result of \cite{review2} to
that in Corollary $1'$:
$$
r=\frac{\delta_k \sqrt{k}t_0}{\tau t_0/3}=\frac{3}{\sqrt{6}+1} (1+(\sqrt{6}+2)\sqrt{k})\delta_k.
$$
It can be concluded from $\delta_k\leq \delta_{k+1}<\frac{1}{\red{1+(\sqrt{6}+2)\sqrt{k}}}$ that
$r<\frac3{\red{\sqrt{6}+1}}<1$, which means that the requirement of $\|\bm b\|_2$ in Corollary $1'$ is more
relaxed.

Second, the requirement of $\delta_{k+1}$ in Corollary $1'$ is more relaxed. Because $\tau$ in
Corollary $1'$ is optional, it can be chosen small enough that
$$
\frac1{\red{1+(\sqrt{6}+2)\sqrt{k}}}<\red{\frac{1-\tau}{\sqrt{k}+1}}.
$$

Despite the difference in requirements, the recovery errors given in Theorem 2 of \cite{review2}
and Corollary $1'$ are the same, since these errors are both derived when the support set of the
sparse signal is perfectly recovered.

\red{\subsection{Corollary 2 and [40, Th.3.1]}}

In \cite{error}, the main result concerns the error estimation for OMP. It is proved that
\begin{align*}
\|\bm x-\textrm{OMP}_S \,\bm x\|_2^2
\red{\leq} 2\|\bm x\|_2 \left(\sigma_S(\bm x)+4\delta_{2S}
(2+\lceil\log_2S\rceil) \|\bm x\|_2\right),
\end{align*}
where $\bm x$ is a non-sparse signal we wish to recover, $\textrm{OMP}_S \,\bm x$ is the estimated
solution via OMP in the $S$th iteration, $\sigma_S(\bm x)$ is the $\ell_2$ error between the best
$S$-term approximation of $\bm x$ and $\bm x$, and $\delta_{2S}$ is the RIC of order $2S$. This
conclusion gives an upper bound on the error between the original signal and the estimated result
of any iteration in OMP.

The original signal to be recovered in \cite{error} is non-sparse, and the inputs $\bm y$ and $\bm
\Phi$ are assumed non-perturbed. Thus the result actually gives an upper bound on the error between
$\bm x$ and $\textrm{OMP}_S \,\bm x$ for ($N_0$) process. Set $S=k$, and this result can be
rewritten as
\begin{equation}\label{r30}
\|\bm x- \hat{\bm x} \|_2^2\leq 2\|\bm x\|_2 \left(\|\bm x^{(2)}\|_2+4\delta_{2k} (2+\lceil\log_2
k\rceil) \|\bm x\|_2\right).
\end{equation}
In Corollary 2, the result is
\begin{equation}\label{r31}
\|\bm x^{(1)}-\hat{\bm x}\|_2 \leq \frac{\sqrt{1+\delta_k}}{\sqrt{1-\delta_k}}(\beta+\gamma)\|\bm
x^{(1)}\|_2.
\end{equation}

Before comparison, it is worth mentioning that there are fundamental differences between the above
two conclusions. First, conditions that guarantee the support set recovery of the best $k$-term
approximation of $\bm x$ is the main concern in Corollary 2, and based on the successful support
recovery, an upper bound on the error is estimated. In the reference, however, the $\ell_2$ error
is directly given regardless of the support recovery. Sometimes, recovering the support set other
than the more accurate estimation is a fundamental concern. Second, compared with \cite{error},
this paper has an apparent limitation: the non-sparse signal considered in this paper is almost
sparse, whereas the one in \cite{error} is arbitrary.

Despite the differences, a tentative comparison of their recovery error estimations is given as
follows. Notice that it is really hard to demonstrate which result is better, since the result in
\cite{error} involves $\delta_{2k}$ which does not appear in our work. However, a condition with
$\delta_{2k}$ involved is given under which (\ref{r31}) is at least as good as (\ref{r30}). From
(\ref{r31}) one has
\begin{align}\nonumber
\|\bm x-\hat{\bm x}\|_2^2 &= \|\bm x^{(1)}-\hat{\bm x}\|_2^2 + \|\bm x^{(2)}\|_2^2 \\ \nonumber
&\leq \frac{1+0.5}{1-0.5}(\beta+\gamma)^2\|\bm x^{(1)}\|_2^2+\|\bm x^{(2)}\|_2^2 \\ \nonumber & =
(3(\beta+\gamma)^2+\beta^2)\|\bm x^{(1)}\|_2^2 \\ \label{r32} &\leq 4(\beta+\gamma)^2 \|\bm
x^{(1)}\|_2^2.
\end{align}
If
\begin{equation}
\delta_{2k} > (\beta+\gamma)^2/4,
\end{equation}
from (\ref{r32}) one has
\begin{equation}\label{r36}
\|\bm x-\hat{\bm x}\|_2^2 \le 16 \delta_{2k} \|\bm x^{(1)}\|_2^2 \leq 16 \delta_{2k} \|\bm x\|_2^2.
\end{equation}
Compared with (\ref{r30}), (\ref{r36}) actually gives a tighter bound.

In fact, the above
requirement of $\delta_{2k}$ can be written in terms of $k$:
\begin{equation}\label{rr1}
\delta_{2k}\geq \red{\frac1{54k}}.
\end{equation}
Assume nontrivially that $\beta+\gamma\neq 0$. Thus $|\textrm{supp}(\bm x^{(1)})|=k$ and $\|\bm
x^{(1)}\|_2\geq \sqrt{k}t_0$. According to (\ref{dj18}) and (\ref{r33}), one has
\begin{equation}\label{r34}
\beta+\gamma \leq \frac{t_0}{3.69\|\bm x^{(1)}\|_2} \leq \frac{1}{3.69\sqrt{k}}.
\end{equation}
Combining (\ref{r32}), (\ref{rr1}) and (\ref{r34}), it holds that
\begin{equation}\label{r35}
\|\bm x-\hat{\bm x}\|_2^2  \le \frac{4}{3.69^2k} \|\bm x^{(1)}\|_2^2 \le16 \delta_{2k} \|\bm
x\|_2^2.
\end{equation}

\red{\subsection{Corollary 4 and [23, Th.4.1]}}

For $(N_0)$ process with $k$-sparse signal $\bm x$, Davenport and Wakin proved
in \cite{Mark} that if $\bm \Phi$ satisfies the RIP of order $k+1$ with $\delta_{k+1} < 1/3,$ and
\begin{equation}\label{r38}
\alpha>  \frac{1+\delta_{k+1}(2\sqrt{k-1}-1)}{1-3 \delta_{k+1}} \triangleq I(\delta_{k+1}),
\end{equation}
then OMP will recover $\bm x$ sequentially from $\bm y$ and $\bm \Phi$ in $k$ iterations
\cite{Mark}.

When $\bm x$ is no longer sparse, and the sensing matrix as well as the
measurement vector is perturbed, Theorem 4 shows that the elements of ${\rm supp}(\bm x)$ can still
be picked up sequentially.

Corollary 4 is derived from Theorem 4. For $k>1$, one has $2\sqrt{k-1}-1\geq 1$. Thus, it can be
seen from (\ref{r29}) and (\ref{r38}) that Corollary 4 is at least as good as the conclusion in
\cite{Mark} when $I(\delta_{k+1})$ is greater than $1.2$ (i.e.
$\delta_{k+1}>(10\sqrt{k-1}+13)^{-1}$), and the latter one is better otherwise.

\section{Conclusion}

In this paper, considering a completely perturbed scenario in the form of $\tilde{\bm y}=\bm \Phi
\bm x+\bm b$ and $\tilde{\bm \Phi}=\bm \Phi+\bm E$, the performance of OMP in recovering an almost
sparse signal (i.e. $\hat{\bm x}=R_{\textrm{OMP}}(\tilde{\bm y}, \tilde{\bm \Phi}, \cdots)$) is
studied.

Though exact recovery of the best $k$-term approximation of $\bm x$ is no longer realistic, Theorem
1 shows that exact recovery of its support via OMP can be guaranteed under suitable conditions.
Based on RIP, such conditions involve the sparsity, the relative perturbations of $\bm y$ and $\bm
\Phi$, and the smallest nonzero entry of $\bm x$. Furthermore, the error between the the best
$k$-term approximation of $\bm x$ and the output $\hat{\bm x}$ is estimated. This completely
perturbed framework extends the prior work in non-perturbed and measurement-perturbed scenarios.
Furthermore, we construct a sensing matrix and perturbations with which an almost sparse signal
cannot be recovered. The RIC of the matrix is slightly bigger than that in the sufficient
conditions of Theorem 1, which indicates that the conditions are rather tight.

In addition, when $\bm x$ is an $\alpha$-strong-decaying signal, several extensions of Theorem 1
are put forward. Theorem 3 reveals that the requirement in Theorem 1 can be relaxed to guarantee
the exact recovery of support. Theorem 4 demonstrates that if $\alpha$ is large enough, the support
is picked up in the order of its entries' magnitude. This advantage is of great significance in
practical scenarios, since the larger entries are often more important than the smaller ones, and
recovery in order indicates the algorithm is more stable. In the end, Theorem 5 discussed the other
scenario of general perturbations, which is in the form of $\tilde{\bm y}=\tilde{\bm \Phi}\bm x+\bm
b$ and $\tilde{\bm \Phi}=\bm \Phi+\bm E$, with the recovery process written as $\hat{\bm
x}=R_{\textrm{OMP}}(\tilde{\bm y}, \bm \Phi, \cdots)$. Notice that several results similar to
Theorem 2-4 are available for this scenario, however, they are not included for simplicity. These
results are in comprehensive comparisons with some previous ones, and conditions under which our
results are at least as good as them are discussed.

\appendix

\appendixtitleon
\begin{appendices}

\section{Proof of Lemma 1}

\begin{proof}
First of all, see $\{x_i~|~i=1,2,\cdots,l-m\}$ as $l-m$ constants, and define the function with
variable $x\ge \alpha x_{l-m}$
\begin{align*}
g(x)=\frac{\red{\sum\limits_{i=1}^{l-m} x_i^2+b_m x^2}}{\left(\sum\limits_{i=1}^{l-m} x_i+c_m x\right)^{2}},
\end{align*}
where $b_m=\sum_{i=0}^{m-1}\alpha^{2i},\,c_m=\sum_{i=0}^{m-1}\alpha^{i},\,1\leq m<l$. Then \red{we prove}
$g(x)\geq g(\alpha x_{l-m})$.

The proof lies in the fact that $g(x)$ can be written as
\begin{align*}
g(x)=\frac{b_m}{c_m^2}-\frac{2\eta}{c_m^2}\frac{y}{(y-\theta+\eta/b_m)^2},
\end{align*}
where
$$
\eta=\frac{b_m}{c_m}\sum_{i=1}^{l-m} x_i,\,\,
\theta=\red{\frac{\frac{b_m}{c_m^2}\left(\sum\limits_{i=1}^{l-m} x_i\right)^2-\sum\limits_{i=1}^{l-m}
x_i^2}{\frac{2b_m}{c_m}\sum\limits_{i=1}^{l-m} x_i}},
$$
and $y=x+\theta$. Because $-\theta+\eta/b_m>0$, $g(x)$ equals its minimum when
\begin{equation}\label{r10}
y^0=-\theta+\eta/b_m,
\end{equation}
which further infers that
\begin{align*}
x^0=y^0-\theta=\eta/b_m-2\theta=\frac{c_m\sum_{i=1}^{l-m} x_i^2}{b_m\sum_{i=1}^{l-m} x_i}.
\end{align*}
Because $x^0\leq x_{l-m}<\alpha x_{l-m}\leq x$ and $g(x)$ is an increasing function when $x\ge
x^0$, $g(x)\geq g(\alpha x_{l-m}).$

Lemma 1 is proved by induction. To begin with, let $m=1$ and fix $\{x_i~|~i=1,\cdots,l-1\}$, then
the above conclusion implies $f(x_1,\cdots,x_{l-1},x_l)\geq f(x_1,\cdots,x_{l-1},\alpha x_{l-1})$.

Furthermore, assume that
\begin{align*}
f(x_1,\cdots,x_{l-m+1},\cdots,x_l)
\red{\geq} f(x_1,\cdots,x_{l-m+1},\cdots,\alpha^{m-1} x_{l-m+1}).
\end{align*}
The above conclusion gives
\begin{align*}
f(x_1,\cdots,x_{l-m},x_{l-m+1},\cdots,\alpha^{m-1} x_{l-m+1})
\geq&f(x_1,\cdots,x_{l-m},x,\cdots,\alpha^{m-1} x)|_{x=\alpha x_{l-m}}\\
=&f(x_1,\cdots,x_{l-m},\cdots,\alpha^m x_{l-m}).
\end{align*}
Therefore, it can be inducted that $f(x_1,\cdots,x_l)$ is no less than $f(x_1,\cdots,\alpha^{l-1}
x_1)$, which concludes the proof.
\end{proof}

\section{Proof of Lemma 2}

\begin{proof}
It can be concluded from (\ref{r16}), (\ref{r17}), and (\ref{r18}) that
\begin{align}
{\bm h}^l &= \tilde{\bm A}_{\Lambda^{l-1}}^\textrm{T} \tilde{\bm P}_{\Lambda^{l-1}}^\bot \tilde{\bm
y} \nonumber\\
&=\red{\tilde{\bm A}_{\Lambda^{l-1}}^\textrm{T}} \tilde{\bm P}_{\Lambda^{l-1}}^\bot (\tilde{\bm
\Phi}\bm x^{(1)}+\bm e)\nonumber\\
&=\tilde{\bm A}_{\Lambda^{l-1}}^\textrm{T} \tilde{\bm A}_{\Lambda^{l-1}} \bm x^{*} + \tilde{\bm
A}_{\Lambda^{l-1}}^\textrm{T} \tilde{\bm P}_{\Lambda^{l-1}}^\bot \bm e\nonumber\\
&=\bm h_1+\bm h_2,\label{r19}
\end{align}
where $\bm h_1=\tilde{\bm A}_{\Lambda^{l-1}}^\textrm{T} \tilde{\bm A}_{\Lambda^{l-1}} \bm x^*$,
$\bm h_2=\tilde{\bm A}_{\Lambda^{l-1}}^\textrm{T} \tilde{\bm P}_{\Lambda^{l-1}}^\bot \bm e$.
Because $\|\bm x^*\|_0+|\Lambda^{l-1}|+1\leq k+1$, according to Lemma 3.3 in \cite{Mark}, for all
$j \notin \Lambda^{l-1} $, it holds that
\begin{equation}\label{r21}
|\bm h_1(j)-\bm x^*(j)|\leq \frac{\tilde{\delta}_{k+1}}{1-\tilde{\delta}_{k+1}} \|\bm x^*\|_2.
\end{equation}
According to Lemma 3.2 in \cite{Mark}, for $j \notin \Lambda^{l-1} $,
\begin{align}
\|\bm h_2(j)\|&= \langle\tilde{\bm A}_{\Lambda^{l-1}} \bm e_j,\tilde{\bm P}_{\Lambda^{l-1}}^\bot
\bm e\rangle \nonumber\\
&\leq  \sqrt{1+\tilde{\delta}_{k+1}} \|\tilde{\bm P}_{\Lambda^{l-1}}^\bot \bm e\|_2\nonumber\\
&\leq \frac{1}{1-\tilde{\delta}_{k+1}} \|\bm e\|_2 \nonumber\\
&\leq
\frac{\varepsilon_h}{1-\tilde{\delta}_{k+1}}.\label{r20}
\end{align}
Notice that the last inequality holds since $\|\bm e\|_2\leq \varepsilon_h$, which has been given
in the proof of Theorem 3. Combining (\ref{r19}), (\ref{r21}), (\ref{r20}), and triangle
inequality, one finally gets
\begin{align*}
|{\bm h}^l(j) - \bm x^*(j)|&\leq |\bm h_1(j)-\bm x^*(j)|+|\bm h_2(j)| \\
&\leq \frac{\tilde{\delta}_{k+1}\|\bm x^*\|_2+\varepsilon_h}{1-\tilde{\delta}_{k+1}}.
\end{align*}
\end{proof}

\end{appendices}

\end{document}